# Quantum Theory on Glucose Transport Across Membrane

Liaofu Luo*

Department of Physics, Inner Mongolia University, Hohhot, 010021 China

## Abstract

After a brief review of the protein folding quantum theory and a short discussion on its experimental evidences the mechanism of glucose transport across membrane is studied from the point of quantum conformational transition. The structural variations among four kinds of conformations of the human glucose transporter GLUT1（ligand free occluded, outward open, ligand bound occluded, and inward open）are looked as the quantum transition. The comparative studies between mechanisms of uniporter (GLUT1) and symporter (XylE and GlcP) are given. The transitional rates are calculated from the fundamental theory. The monosaccharide transport kinetics is proposed. The steady state of the transporter is found and its stability is studied. The glucose (xylose) translocation rates in two directions and in different steps are compared. The mean transport time in a cycle is calculated and based on it the comparison of the transport times between GLUT1,GlcP and XylE can be drawn. The non-Arrhenius temperature dependence of the transition rate and the mean transport time is predicted. It is suggested that the direct measurement of temperature dependence is a useful tool for deeply understanding the transmembrane transport mechanism.

*Email : lolfcm@imu.edu.cn

Recently, the crystal structures of several bacterial and human monosaccharide transporters were reported [1-4]. Through sequential and structural comparison with other members of the sugar porter subfamily, the basic transport mechanism of the human glucose GLUT1 is clarified [4]. It was proposed that the successive conformational changes of the transporter occur in the glucose transport process and form a complete cycle, from ligand free occluded conformation (A), changed to outward open (B), ligand bound occluded (C), and inward open (D), then to the ligand free occluded of the next cycle. The conformation A is connected to the intracellular side and the conformation C to the extracellular side. The above picture provides a basis for understanding the general transport dynamics for sugar porter subfamily. However, more detailed and quantitative analysis is necessary. Since the glucose transport is essentially a process associated with a series of conformational changes of the porter protein we shall discuss the problem by using the quantum theory of protein conformation transition which was developed in recent years by the author [5][6][7].

# 1 Quantum transition between conformational states

## 1.1 Conformational change as torsion transition

Considering that the torsion vibration energy 0.03-0.003 ev is the lowest in all forms of biological energies, even lower than the average thermal energy per atom at room temperature and easily changed at physiological temperature, we can look upon the torsion as the slow variable of the macromolecule. Following Haken's synergetics, the slow variables always slave the fast ones of a complex system. By taking a general form of the torsion Hamiltonian $H_1$ and the fast variable Hamiltonian $H_2$ and by the adiabatically elimination of fast variables we obtained a Hamiltonian describing the conformational transition of the macromolecule and deduced a general formula for the conformational transition rate [5].

$$W = \frac{2p}{\mathbf{h}^2 \bar{w}'} I'_V I'_E$$

$$I'_V = \frac{\mathbf{h}}{\sqrt{2p} dq} \exp\{\frac{\Delta G}{2k_B T}\} \exp\{\frac{-(\Delta G)^2}{2\bar{w}^2 (dq)^2 k_B T \sum_j^N I_j}\} (k_B T)^{1/2} (\sum_j^N I_j)^{-1/2} \quad (1)$$

$$I'_E = \sum_j^M |a_{a'a}^{(j)}|^2 \cong M \bar{a}^2$$

where $I'_V$ is slow-variable (molecular torsion) factor, $\Delta G$ is the free energy difference between initial and final states, $N$ is the number of torsion modes participating in the quantum transition, $I_j$ is the inertia moment of the $j$-th torsion mode ($I_0$-their average), $\bar{w}$ $\bar{w}'$ and $dq$ are torsion potential parameter, $I'_E$ is fast-variable factor, $M$ — the number of torsion modes coupling to fast variables, and $\bar{a}^2$ — the matrix element square of fast-variable Hamiltonian $H_2$, expressed as:

$$\bar{a}^2 = \frac{\mathbf{h}^2}{I_0} \left| \frac{e_{a'a}}{e_a^{(0)} - e_{a'}^{(0)}} \right|^2 \quad (a' \neq a)$$

$$e_{a'a} = \langle e_{a'a}^{(j)} \rangle_{av} \qquad e_{a'a}^{(j)} = \langle a' | (\frac{\partial H_2}{\partial q_j})_0 | a \rangle \quad (2)$$

($e_a^{(0)}$ and $e_{a'}^{(0)}$ are eigenvalues of $H_2$ at $\{q_j = q_j^{(0)}\}$). The rate formula Eq (1) has been successfully applied to the protein folding [5][6][7]. It explained the curious non-Arrhenius temperature-dependencies of the folding rate for all proteins whose experimental data were reported. It also explained the specific statistical distribution of the folding rates for all measured two-state proteins. The correlation between theoretical prediction with experimental folding rate attained 73%–78% [8]. The multi-state protein folding can also be regarded as quantum conformational transitions similar to two-state proteins but with an intermediate delay [9]. Therefore, it is reasonable to assume the above equations (1) (2) can be used for membrane protein and provide a basis for studying the transport dynamics of glucose across the membrane.

## 1.2 Applicability of quantum theory in macromolecule

About the applicability of quantum theory in macromolecule a fundamental problem is how to estimate the decoherence effect for the system. If the decoherence effect is strong enough then the quantum picture would cease to be effective for a macromolecule. The decoherence effect is estimated by computing the decoherence time of the system under thermal environment. The rigorous solution of decoherence time is difficult but some simple models were proposed. One such model introduced by Zurek in ref. [10][11] showed the decoherence time

$$t_D = t_R (\frac{h}{\Delta x \sqrt{2mk_B T}})^2$$

where $t_R$ is the relaxation time due to the interaction of the particle with a scalar field and $\frac{h}{\sqrt{2mk_B T}}$ the thermal de Broglie wavelength, $m$ the particle mass and $\Delta x$ the dimension of particle. It leads to $t_D = 10^2 t_R$ for an electron but about $10^{-2} t_R$ for an atom (carbon), much shorter than the electronic decoherence time. However, from the idea of quantum correlation relativity the coherence in different degrees of freedom of a macromolecule can exhibit differently [12]. While the center-of-mass degrees of freedom are influenced by the environment the other degrees of freedom (subsystems) can still maintain their quantum nature. To study the problem we use Zurek's model to the torsional angular motion in protein folding. It is easily to deduce a similar formulas for torsional decoherence time:

$$t_D^{(tor)} = t_R (\frac{h}{\Delta q \sqrt{2Ik_B T}})^2 > 10^4 t_R (\frac{h}{\sqrt{2Ik_B T}})^2 \cong t_R$$

（3）

where $\Delta q$ is the uncertainty of torsional angle, $I$ the torsional inertia moment of atomic group and $\sqrt{2Ik_B T} = J_{therm}$ the thermal angular momentum. By using the uncertainty relation and $\Delta q\, \Delta J \approx \hbar$ ($\Delta J$ is the uncertainty of angular momentum) the first equality of eq. (3) can be rewritten as $t_D^{(tor)} \approx t_R (\frac{\Delta J}{J_{therm}})^2$. In the next inequality of eq. (3), $\Delta q \leq 0.5$ degree (about one tenth of the angular shift $dq$ in torsion potential) and $I = 10^{-37} \text{g} \cdot \text{cm}^2$ (for a typical atomic group in polypeptide) have been taken. By comparing the torsion decoherence with electronic and atomic decoherence time we find, if the relaxation rates ($\frac{1}{t_R}$) are same in three cases then the decoherence effect on molecular torsion is in the midst of the electron and the atom. In fact, because of the thermal average over torsional vibration states used in theoretical calculation the thermal excitation effect of the scalar field (representing the environmental perturbation) has been partly taken into account. The surplus interaction of the scalar field interaction with torsion subsystem should be weaker and the torsional relaxation rate decreases. Thus, even if the quantum coherence for the macromolecule as a single particle may have been destroyed the coherence in the torsional degree of freedom still works.

About the applicability of quantum theory in macromolecule we should notice that the quantum coherence in cytoskeletal microtubules and its importance for neuron activity and brain function have been widely discussed by many authors [13]-[16]. On the other hand we indicate that the multi-torsion correlation has observed in the two-state protein folding which give further supports on the existence of the torsional quantum coherence in macromolecule [5].

More direct experimental evidences on the quantum nature of protein folding can be found as follows. Set the folding rate $1/\tau$ for a two-state protein. In classical theory, the torsion-angle variations $\Delta_t j_i$, $\Delta_t y_i$ and $\Delta_t c_i$ at time $t<\tau$ should be condensed to a value near 0, since in the duration shorter than the average folding time $\tau$ the conformational change of the two-state protein has not occurred. But in quantum theory, though $t<\tau$, the time duration from 0 to $t$ is enough for finding a torsional jump since the quantum transition always occurs instantaneously. In fact, the quantum jump occurs stochastically at any instant within the folding time $\tau$. Qiu et al [17] used laser temperature-jump spectroscopy to measure the unfolding rate of 20-residue Trp-cage protein. In their experiment the resulting T-jump of 5-20$^o$ C occurs within 20-30 ns and the thermal unfolding enhance the Trp emission. They discovered the Fluorenscence Intensity (FI) increases rapidly from 11.5mV to 14 mv in a time duration of 4μs, and therefore determined the unfolding rate 4μs. However, if the unfolding of the two-state Trp-cage protein obeys classical law the fluorescence should emit only after the unfolding has been closed up. So the fluorenscence enhance would be observed with a time delay of 4μs after the initial drop owing to the intrinsic T dependence of Trp emission. But no such delay was observed in Qiu's experiment. Rather, the unfolding event occurred stochastically in the time duration of 4μs. Therefore, one may conclude that the observed FI – t relation is in accordance with the quantum explanation on the unfolding of the Trp-cage protein. It is expected that the further observation of instantaneous dihedral transition in the timescale of microseconds will be able to give more evidences on the quantum nature of protein folding.

## 2  Kinetic model for transporter

## 2.1  Equations for sugar transport across membrane

The human glucose transporter GLUT1 works in the following cycle [4]:

(1)  $A \rightleftarrows B$

(2)  $B + g_{out} \rightleftarrows C$

(3)  $C \rightleftarrows D$

(4)  $D \rightleftarrows A + g_{in}$

with rate $k_i$ for the *i*-th step and $k_i'$ for the *i*-th reverse step (*i*=1,2,3,4) of the reaction. Here A=(ligand free occluded), B= (ligand free and outward open), C= (ligand bound occluded) and D=(ligand bound and inward open) represent four conformations of GLUT1, $g_{out}$ and $g_{in}$ represent the extracellular or intracellular glucose molecule respectively (Fig 1). Following mass-action kinetics we write the concentration equations for GLUT1 in four structural states and for the glucose in two states. Set the concentrations

[A]=*x*,  [B]=*y*,  [C]=*z*,  [D]=*w*,  [$g_{out}$]=*v*,  [$g_{in}$]=*u*

The reaction equations read

$$\frac{dv}{dt} = s + k_2' z - k_2 vy$$

$$\frac{du}{dt} = -ru + k_4 w - k_4' ux$$

$$\frac{dx}{dt} = k_1' y - k_1 x + k_4 w - k_4' xu$$

$$\frac{dy}{dt} = k_1 x - k_1' y + k_2' z - k_2 vy \tag{4}$$

$$\frac{dz}{dt} = k_2 vy - k_2' z + k_3' w - k_3 z$$

$$\frac{dw}{dt} = k_3 z - k_3' w - k_4 w + k_4' ux$$

In writing Eq (4) the following boundary conditions have been assumed,

$$(\text{initial reservoir chemicals}) \xrightarrow{s} g_{out} \text{ with constant rate } s,$$

$$g_{in} \xrightarrow{r} (\text{final reservoir chemicals}).$$

From Eq (4) one obtains

$$x + y + z + w = C \tag{5}$$

immediately. So the variable $w$ can be eliminated and the independent variables of the system are only $x, y, z, u$ and $v$. The rate constants $k_i$ and $k_i'$ ($i$=1,2,3,4) in equations (4) should be calculated from quantum theory of conformational transition.

The above model for GLUT1 can be generalized to bacterial monosaccharide transporters GlcP and XylE [2][3]. For glucose transporter GlcP, by assuming the proton–binding to decrease the energetic barrier mediating different conformations a 6-state model was suggested in literature [3]: A=outward facing, unliganded; B=outward facing, $H^+$-bound; C=outward facing, substrate-bound; D=inward facing, substrate-bound; E= inward facing, $H^+$-bound but glucose-released; F= inward facing, unliganded (both $H^+$ and glucose released). The transition rates between neighboring two states in a cycle are denoted as $k_i$ and $k_i'$ respectively ($i$=1,2,3,4,5,6) (Fig 2). Set the concentration of six states is denoted by x, y, z, z', y' or x' for A, B, C, D, E, F respectively. The inside/outside glucose concentration is denoted by $u$ and $v$ as before. The $H^+$ concentration is supposed as constant. The concentration equations can be written similarly as Eq (4), namely

$$\frac{dx}{dt} = k_1' y - k_1 x c_p + k_6 x' - k_6' x$$

$$\frac{dy}{dt} = k_1 x c_p - k_1' y + k_2' z - k_2 v y$$

$$\frac{dz}{dt} = k_2 v y - k_2' z + k_3' z' - k_3 z$$

$$\frac{dz'}{dt} = k_3 z - k_3' z' + k_4' u y' - k_4 z'$$

$$\frac{dy'}{dt} = k_4 z' - k_4' u y' + k_5' c_p x' - k_5 y'$$

$$\frac{dx'}{dt} = k_5 y' - k_5' c_p x' + k_6' x - k_6 x'$$

$$\frac{dv}{dt} = s + k_2' z - k_2 v y$$

$$\frac{du}{dt} = -ru + k_4 z' - k_4' u y' \tag{6}$$

When the lifetime of the state B and E is short enough the steps of $A + H^+ \to B$ and $B + g_{out} \to C$ are merged into one step $A + H^+ + g_{0ut} \to C$ and $D \to g_{in} + E$ and $E \to H^+ + F$ are merged into $D \to g_{in} + H^+ + F$. The 6-state model is degenerated to the 4-state model.

From Eq (6) one has

$$x + y + z + z' + y' + x' = C \tag{7}$$

similar to Eq (5).

While GLUT1 is a uniporter that catalyses the translocation of glucose down its concentration gradient across the membrane the bacterial transporter GlcP or XylE is a proton symporter that exploits the transmembrane proton gradient to drive the "uphill" translocation of substrate against its concentration gradient. The above 6-state model for transporter GlcP can be used for xylose transport of XylE as well. However, considering that XylE has several distinct features in comparison with other sugar:$H^+$ symporters, such as a large number of hydrogen-bond-forming residues (that is related to abundant protonation and deprotonation) distributed on the interface between the transmembrane segments and the intracellular domain [1] we propose an alternative mechanism for the proton symporter XylE. Suppose there exist three conformational states for the transporter XylE: ligand free state α, ligand and proton bound state $β_p$ and ligand bound state β. After the xylose molecules producted in the process of "reservoir $\to g_{out}$" with constant rate $s$, the conformation of XylE changes subsequently in three steps (Fig 3):

(1) $g_{out} + a + p \rightleftharpoons b_p$    with rate $k_a$ and $k_a$' respectively

(2) $b_p \to b + p$    with rate $k_b$

(3) $b \rightleftharpoons a + g_{in}$    with rate $k_c$ and $k_c$' respectively

Finally, the released xylose $g_{in}$ is exploited through the process "$g_{in} \xrightarrow{r}$ final reservoir chemicals". In the above-mentioned 3-step reaction the first step describes the ligands (xylose and proton) binding where the proton concentration is assumed to be a constant, $c_p$. The second step describes the deprotonation process of the symporter by which the released proton is assumed quickly absorbed to the medium and the reverse of the process does not happen actually. It is assumed that only the deprotonated symporter can undergo the conformation switch from inward to outward and release the substrate. This is described by step 3. The essential point of the model is the assumption of only three states α, β and $β_p$ existing in the conformational transition, corresponding to the observed three states: inward open, partially occluded inward open and outward-facing in XylE [2]. Set the transporter and xylose concentrations

[α]=$x$,  [$β_p$]=$y$,  [β]=$z$,  [$g_{out}$]=$v$,  [$g_{in}$]=$u$

The reaction equations are

$$\frac{dv}{dt} = s + k_a' y - k_a c_p vx$$

$$\frac{du}{dt} = -ru + k_c z - k_c' ux$$

$$\frac{dx}{dt} = k_a' y - k_a c_p vx + k_c z - k_c' ux \qquad (8)$$

$$\frac{dy}{dt} = k_a c_p vx - k_a' y - k_b y$$

$$\frac{dz}{dt} = k_b y - k_c z + k_c' ux$$

From $\frac{dx}{dt} + \frac{dy}{dt} + \frac{dz}{dt} = 0$, it leads to

$$x + y + z = C \qquad (9)$$

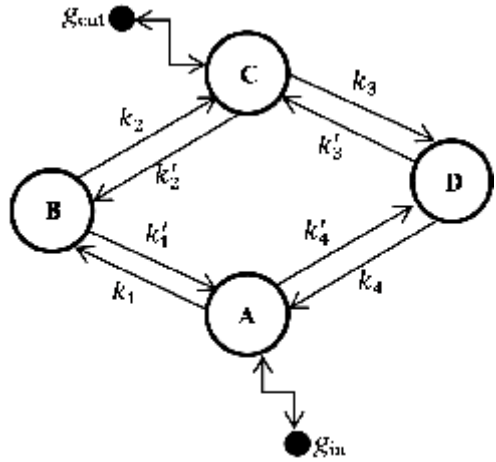

Figure 1　Quantum transition between conformational states of human glucose transporter GLUT1.　A=ligand-free occluded，B=outward open，C=ligand-bound occluded，D=inward open.

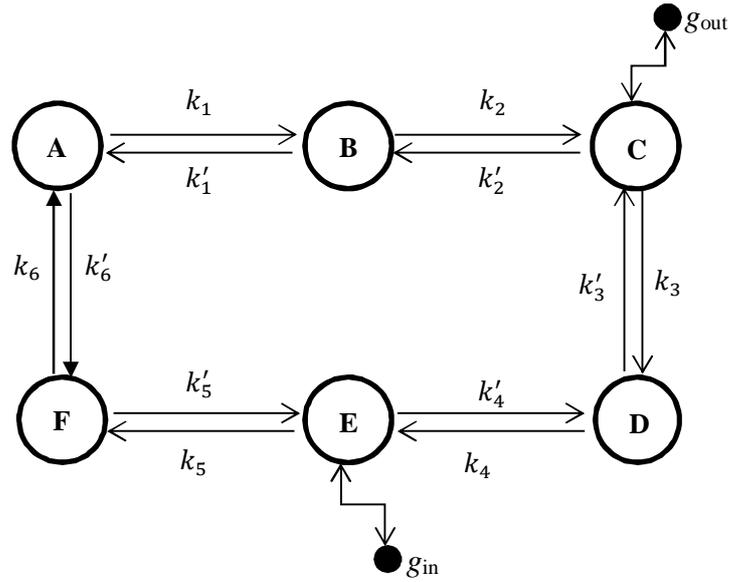

Figure 2　Quantum transition between conformational states of bacterial glucose/H$^+$ symporter GlcP.　A=outward facing, unliganded; B=outward facing, H$^+$-bound; C=outward facing, glucose-bound; D=inward facing, substrate-bound; E=inward facing, glucose-released; F=inward facing, unliganded (both H$^+$ and glucose released)

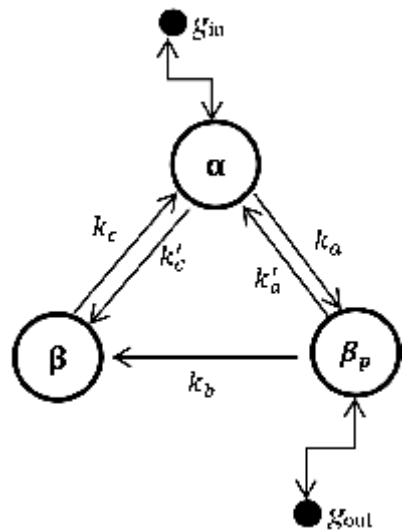

Figure 3　Quantum transition between conformational states of bacterial transporter XylE.　α = ligand-free state,　β$_p$ = ligand- and proton-bound state,　β = ligand-bound (proton-free) state.

## 2.2 Steady state and stability

We shall discuss the steady state of the transporter and its stability. For the four-state model of GLUT1, by setting the left-handed-site of Eq (4) being zero we obtain steady state solution. Because of

$$\frac{dx}{dt} + \frac{dy}{dt} = \frac{dv}{dt} + \frac{du}{dt} + ru - s \tag{10}$$

It leads to the steady state of $u$,

$$u_0 = \frac{s}{r} \tag{11}$$

Through direct calculations we obtain the steady state of other concentrations $x_0$, $y_0$, $z_0$ and $v_0$

$$y_0 = \frac{k_1 x_0 - s}{k_1'} \tag{12}$$

$$(k_3' + k_3)z_0 = -k_3'(1 + \frac{k_1}{k_1'})x_0 + (1 + \frac{k_3'}{k_1'})s + k_3'C \tag{13}$$

$$v_0 = \frac{k_2' z_0 + s}{k_2 y_0} = (\frac{k_2' z_0 + s}{k_1 x_0 - s}) \frac{k_1'}{k_2} \tag{14}$$

For given $x_0 > \frac{s}{k_1}$, one obtains $y_0$, $z_0$, and $v_0$ readily from (12)(13) and (14). The concentrations $x_0$ cannot be fully determined that means more than one sets of steady states existing.

The liner equations near a given set of steady states ($v_0$, $u_0$, $x_0$, $y_0$, $z_0$) are deduced from Eq (4) as

$$\frac{ddv}{dt} = k_2' dz - k_2 v_0 dy - k_2 y_0 dv$$

$$\frac{ddu}{dt} = -(k_4 + k_4' u_0)dx - k_4 dy - k_4 dz - k_4' x_0 du - rdu$$

$$\frac{ddx}{dt} = -(k_1 + k_4 + k_4' u_0)dx + (k_1' - k_4)dy - k_4 dz - k_4' x_0 du$$

$$\frac{ddy}{dt} = k_1 dx - (k_1' + k_2 v_0)dy + k_2' dz - k_2 y_0 dv$$

$$\frac{ddz}{dt} = -k_3' dx + (k_2 v_0 - k_3')dy - (k_2' + k_3 + k_3')dz + k_2 y_0 dv \tag{15}$$

Denote the coefficient matrix of the above equations as $M$, we have

$$M - lI =$$

$$\begin{bmatrix} -k_2 y_0 - \lambda & 0 & 0 & -k_2 v_0 & k_2^{'} \\ 0 & -(k_4^{'} x_0 + r) - l & -(k_4 + k_4^{'} u_0) & -k_4 & -k_4 \\ 0 & -k_4^{'} x_0 & -(k_1 + k_4 + k_4^{'} u_0) - \lambda & k_1^{'} - k_4 & -k_4 \\ -k_2 y_0 & 0 & k_1 & -(k_1^{'} + k_2 v_0) - \lambda & k_2^{'} \\ k_2 y_0 & 0 & -k_3^{'} & k_2 v_0 - k_3^{'} & -(k_2^{'} + k_3 + k_3^{'}) - \lambda \end{bmatrix}$$

(16)

From $|M - lI| = 0$ one solves $l_1$, $l_2$, $l_3$, $l_4$ and $l_5$ and obtains

$$l_1 + l_2 + l_3 + l_4 + l_5 = -(k_1 + k_1^{'} + k_2 v_0 + k_2 y_0 + k_2^{'} + k_3 + k_3^{'} + k_4 + k_4^{'} x_0 + k_4^{'} u_0 + r)$$

(17)

Evidently, equation (17) is smaller than 0; so the glucose transport across membrane is globally stable.

From fluctuation matrix (16) we find that the fluctuation of glucose output $du$ is mainly related to the concentration variation $dx$ of GLUT1, while the fluctuation of glucose input $dv$ is mainly related to the concentration variation $dy$ and $dz$. If the input and the output of the glucose have been given at some steady state then the fluctuation of molecular concentration $x$, $y$, $z$ takes a form of $a_1 \exp l_1 t + a_2 \exp l_2 t + a_3 \exp l_3 t$. The relaxing time of three modes is determined by the 3×3 submatrix of the matrix (16). One has

$$l_1 + l_2 + l_3 = -(k_1 + k_1^{'} + k_2 v_0 + k_2^{'} + k_3 + k_3^{'} + k_4 + k_4^{'} u_0)$$

and if $k_2^{'} \ll k_2 v_0$, $k_4 \ll k_4^{'} u_0$ (see next section),

$$l_1 = -(k_3 + k_3{'} + k_2{'})$$

$$l_2 = -\frac{k_1 + k_1^{'} + k_4^{'} u_0 + k_2 v_0}{2} - \frac{|k_1 - k_1^{'} + k_4^{'} u_0 - k_2 v_0|}{2}$$

$$l_3 = -\frac{k_1 + k_1^{'} + k_4^{'} u_0 + k_2 v_0}{2} + \frac{|k_1 - k_1^{'} + k_4^{'} u_0 - k_2 v_0|}{2} \quad (18)$$

All eigenvalues $l_1$, $l_2$ and $l_3$ smaller than 0 show again the stability of the steady state.

The above discussions can easily be generalized to glucose transporter GlcP. Particularly, from the 6-state model Eq (6), it leads to

$$\frac{dv}{dt} + \frac{du}{dt} + ru - s = -(\frac{dz}{dt} + \frac{dz^{'}}{dt}) \qquad (19)$$

and the steady state condition Eq (11) holds in the case, too. The linear equations near steady states give the eigenvalues $l_1$ to $l_7$ and it can be proved

$$\sum_i^7 l_i = -(k_1 c_p + k_1' + k_2 y_0 + k_2 v_0 + k_2' + k_3 + k_3' + k_4 + k_4' u_0 \\ + k_4' y_0 + k_5 + k_5' c_p + k_6 + k_6' + r) < 0 \tag{20}$$

It means the glucose transport in GlcP is also globally stable.

For the bacterial symporter XylE, from the three-state model Eq (8) one has

$$\frac{du}{dt} + \frac{dv}{dt} - \frac{dx}{dt} = s - ru \tag{21}$$

It also leads to the steady state Eq (11) $u_0 = \frac{s}{r}$. Further, from Eq (8) we obtain the steady states of the symporter XylE $y_0$, $x_0$, and $v_0$ as follows.

$$y_0 = \frac{s}{k_b} \tag{22}$$

$$x_0 = \frac{k_c C - (1 + \frac{k_c}{k_b})s}{k_c + k_c' \frac{s}{r}} \tag{23}$$

$$c_p v_0 x_0 = \frac{(k_b + k_a')s}{k_a k_b} \tag{24}$$

Different from GLUT1, under given $s$, $r$ and $C$ only one steady state for XylE can be attained as $k_c C > (1 + \frac{k_c}{k_b})s$.

By the linearization of Eq (8) we deduce the eigenvalues $l_1$ to $l_4$ and obtain the stability condition

$$\sum_i^4 l_i = -(k_a c_p v_0 + k_c' u_0 + k_c + k_a' + k_b + k_a c_p x_0 + k_c' x_0 + r) < 0 \tag{25}$$

# 3 Dynamical aspects of sugar transport

## 3.1 Rate calculated from quantum theory

The rate constants $k_i$, $k_i'$ are key parameters in the theory, we shall use quantum transition theory to deduce them. For transporter GLUT1 all rates

$$K_1 = k_1 = W(A \to B), \quad K_2 = k_2 v_0 = W(B \to C), \quad K_3 = k_3 = W(C \to D), \quad K_4 = k_4 = W(D \to A)$$

and their reverses

$$K_1' = k_1' = W(B \to A), \quad K_2' = k_2' = W(C \to B), \quad K_3' = k_3' = W(D \to C), \quad K_4' = k_4' u_0 = W(A \to D)$$

are in the dimension of (1/time) and can be calculated from Eq (1).

The glucose molecules are hydrogen-bonded to several residues（for example, Gln，Asn，Trp）of the transporter GLUT1. Set the binding energy of one sugar molecule being $E_b$. The N- and C- domain of GLUT1 are connected by an intracellular helical bundle (ICH) which comprises four short helices. The ICH domain serves as a latch that tightens the intracellular gate. The conformational change between inward-open (or outward-open) and occluded is related to the rigid-body rotation of the N and C domains which may be achieved through the bond-length and bong-angle variation of ICH residues [4]. On the other hand, the transmembrane segments TM1 and TM7 contacts each other and constitutes an extracellular gate. The observed conformational change is related to the contact variation of a few residues of the transmembrane segments [4]. It is reasonable to assume there exist two or more minima in the bond-length and bong-angle potential and the conformation change occurs through the quantum transition among them, jumping from one minimum to another. Set the conformational energy of the outward open（or inward open）relative to the occluded denoted by $E_p$ (neglecting the difference between the outward open and inward open).

Thus, we have an approximate estimate of the free energy difference $\Delta G^{(i)}$ for the $i$-th step quantum transition,

$$\Delta G^{(1)} = -E_p, \quad \Delta G^{(2)} = E_p + E_b, \quad \Delta G^{(3)} = -E_p, \quad \Delta G^{(4)} = E_p - E_b \quad (26)$$

Both the ligand binding and the bond-length and bond-angle variation are fast-variables. The fast variables are coupled with many torsion modes. Suppose the number of torsion modes participating in the $i$-th quantum transition denoted by $N_i$ and all $N_i$ modes coupled with fast variables. According to the estimate of the number of residues that are related to the ligand binding or to the bond stretching-bending one may assume $M_i = N_i \approx 200$ or more in each of the four steps of quantum transition.

From Eqs (1) and (2) the rate constant $K_i$ is given by

$$\ln K_i = \frac{\Delta G^{(i)}}{2k_B T} - \frac{(\Delta G^{(i)})^2}{2k_B T N_i J_0} + \frac{1}{2}\ln N_i + \frac{1}{2}\ln T + \ln \bar{a}_i^2 + const \quad (27)$$

with $J_0 = \bar{w}^2 (dq)^2 I_0$, $I_0$–the inertia moment of typical atomic group, $I_0 = 10^{-37} g \cdot cm^2$, $dq\bar{w} \approx 10^{11} s^{-1}$ taken from the protein folding data [5][8].

The reverse rate constants $K_i'$ are related to $K_i$ as [5]:

$$\ln\{\frac{K_i}{K_i^{'}}\} = \frac{\Delta G^{(i)}}{k_B T} + \frac{(\Delta G^{(i)})^2}{2k_B T e}(\frac{\overline{w}^2 - \overline{w}^{'2}}{\overline{w}^{'2}}) + \ln\frac{\overline{w}}{\overline{w}'} \cong \frac{\Delta G^{(i)}}{k_B T}$$ （28）

which can be deduced directly from Eq (1). Here the initial torsion parameter $\overline{w}$ near to the final $\overline{w}'$ has been assumed [8].

Since both glucose binding energy $E_b$ (typically several ev) and conformational energy $E_p$ (typically 0.13ev for one stretching-bending degree of freedom) are much larger than $k_B T$ (0.026ev at room temperature) we have

$$K_2^{'} \ll K_2 \qquad K_1^{'} \gg K_1 \qquad K_3^{'} \gg K_3 \qquad (29)$$

The rate $K_4$ is smaller than $K_4^{'}$ if $E_p < E_b$ or larger than $K_4^{'}$ if $E_p > E_b$. The latter case occurs as the conformation energy comes from the contribution of the sum of many vibration modes. Eq (29) means for GLUT1 the reverse transition from the conformation C to B can be neglected as compared with that from B to C, while the reverse transitions from A to B and from C to D are important in the glucose transport cycle across membrane.

For glucose transporter GlcP, the free energy difference $\Delta G^{(i)}$ in the $i$-th step ($i=1,\ldots,6$) quantum transition can be parameterized as

$$\Delta G^{(1)} = E_H, \qquad \Delta G^{(2)} = E_B, \qquad \Delta G^{(3)} = -E_p,$$
$$\Delta G^{(4)} = -E_B, \qquad \Delta G^{(5)} = -E_H, \qquad \Delta G^{(6)} = E_p \qquad (30)$$

where both $E_B$ and $E_H$ include substrate (glucose or $H^+$) binding energy and conformation-changing energy, $E_p$ is the conformation energy between out-facing and inward-facing. We have

$$K_i^{'} \ll K_i \quad (i=1,2,6) \qquad K_i^{'} \gg K_i \quad (i=3,4,5) \qquad (31)$$

where $K_i = k_i$ for $i=3,4,5,6$ and $K_1 = k_1 c_p$, $K_2 = k_2 v_0$; and $K_i' = k_i'$ for $i=1,2,3,6$ and $K_5^{'} = k_5^{'} c_p$, $K_4^{'} = k_4^{'} u_0$

For bacterial symporter XylE, the rates $K_a = k_a c_p v_0$, $K_b = k_b$, $K_c = k_c$ and $K_a^{'} = k_a^{'}$, $K_c^{'} = k_c^{'} u_0$ can also be found from quantum theory and they obey Eqs (27) and (28), too. Instead of (26) and (30) the free energy difference $\Delta G^{(i)}$ can be parameterized as

$$\Delta G^{(a)} = E_b + dE_b, \qquad \Delta G^{(b)} = -dE_b, \qquad \Delta G^{(c)} = -E_b \qquad (32)$$

($E_b$ - xylose binding and related conformational change energy and $dE_b$ - protonation energy). One has

$$K_a >> K_a' \qquad K_c << K_c' \qquad (33)$$

## 3.2 Mean transport time

The total time needed for glucose transport across membrane in a cycle is called the mean transport time of glucose. For GLUT1 the mean transport time is

$$t = \frac{1}{K_1} + \frac{1}{K_2} + \frac{1}{K_3} + \frac{1}{K_4} = \frac{1}{K_2}(1 + \frac{K_2}{K_3} + \frac{K_2}{K_4} + \frac{K_2}{K_1}) \qquad (34a)$$

(for transport in positive direction or the positive transport) or

$$t' = \frac{1}{K_1'} + \frac{1}{K_2'} + \frac{1}{K_3'} + \frac{1}{K_4'} = \frac{1}{K_2'}(1 + \frac{K_2'}{K_3'} + \frac{K_2'}{K_4'} + \frac{K_2'}{K_1'}) \qquad (34b)$$

(for the negative transport). Notice the following relations between rate constants hold,

$$\ln \frac{K_2}{K_3} = \frac{(2E_p + E_b)}{2k_B T}(1 - \frac{E_b}{NJ_0}) + \ln \frac{\bar{a}_2^2}{\bar{a}_3^2}$$

$$\ln \frac{K_2}{K_4} = \frac{E_b}{k_B T}(1 - \frac{2E_p}{NJ_0}) + \ln \frac{\bar{a}_2^2}{\bar{a}_4^2}$$

$$\ln \frac{K_2}{K_1} = \frac{(2E_p + E_b)}{2k_B T}(1 - \frac{E_b}{NJ_0}) + \ln \frac{\bar{a}_2^2}{\bar{a}_1^2}$$

$$\ln \frac{K_2'}{K_3'} = -\frac{(2E_p + E_b)}{2k_B T}(1 + \frac{E_b}{NJ_0}) + \ln \frac{\bar{a}_2^2}{\bar{a}_3^2}$$

$$\ln \frac{K_2'}{K_4'} = -\frac{E_b}{k_B T}(1 + \frac{2E_p}{NJ_0}) + \ln \frac{\bar{a}_2^2}{\bar{a}_4^2}$$

$$\ln \frac{K_2'}{K_1'} = -\frac{(2E_p + E_b)}{2k_B T}(1 + \frac{E_b}{NJ_0}) + \ln \frac{\bar{a}_2^2}{\bar{a}_1^2} \qquad (35)$$

Taking both $E_b$ and $2E_p$ much larger than $NJ_0 (=2\times 10^{-13}$ erg) and $k_B T$ into account, and estimating the difference between $\bar{a}_1^2$ (or $\bar{a}_3^2$, $\bar{a}_4^2$) and $\bar{a}_2^2$ by a factor no larger than $10^2$, we obtain $\frac{K_2}{K_3} <<1$, $\frac{K_2}{K_4} <<1$, $\frac{K_2}{K_1} <<1$, $\frac{K_2'}{K_3'} <<1$, $\frac{K_2'}{K_4'} <<1$ and $\frac{K_2'}{K_1'} <<1$. Therefore,

$$t = \frac{1}{K_2}$$

$$t' = \frac{1}{K_2'} \tag{36}$$

The above calculation shows that the glucose transport time in a cycle is determined mainly by the step of $B + g_{out} \leftrightarrows C$ (by the rate $K_2$ or $K_2'$). Moreover, the positive transport time is smaller than the negative due to $K_2' \ll K_2$. This means the net transport of glucose is in the direction from extracellular to intracellular.

For GlcP the same argument gives

$$t = \sum_i \frac{1}{K_i} = \frac{1}{K_4} \qquad t' = \sum_i \frac{1}{K_i'} = \frac{1}{K_2'} \tag{37}$$

as $E_B$ is the largest among $E_p$, $E_H$ and $E_B$.

For bacterial symporter XylE, the xylose transports from the ligand free state $\alpha$ across the intermediate $\beta_p$, then to the ligand bound state $\beta$, only in one way and one direction. The mean transport time is

$$t = \frac{1}{K_a} + \frac{1}{K_b} + \frac{1}{K_c} \approx \frac{1}{K_a} \quad (dE_b > NJ_0)$$

$$\approx \frac{1}{K_c} \quad (dE_b < NJ_0) \tag{38}$$

For the protonation energy $dE_b > NJ_0$ the transport time is determined mainly by the transition $g_{out} + a + p \to b_p$; for $dE_b < NJ_0$ the transport time is determined mainly by the transition $b \to a + g_{in}$.

It is interesting to note that there exist some relations on the mean transport time between GLUT1 GlcP and XylE in the present theory. For example, from Eqs (36) (38) and (27) we have an approximate expression

$$\ln \frac{t(\text{XylE})}{t(\text{Glut})} = \ln \frac{K_2}{K_c}$$
$$\cong \frac{\Delta G^{(2)} - \Delta G^{(c)}}{2k_B T} - \frac{1}{2k_B T J_0} \left( \frac{(\Delta G^{(2)})^2}{N_{Glut}} - \frac{(\Delta G^{(c)})^2}{N_{XylE}} \right) \tag{39}$$

for the case of $dE_b < N_{XylE} J_0$ where $\Delta G^{(2)}$ and $\Delta G^{(c)}$ are given by (26) and (32) respectively and the difference of $N$ between GLUT1 and XylE has been marked in Eq (39). For the case of $dE_b > N_{XylE} J_0$ similar results can be obtained only if $\Delta G^{(c)}$ in (39) is replaced by $\Delta G^{(a)}$.

## 3.3 Temperature dependence of transport time

The rate of conformational transition for biological macromolecule is temperature dependent. The relation has non-Arrhenius peculiarity on the plot of logarithm rate versus $1/T$ and that was proved in protein folding [5][8]. For monosaccharide transport we predict the similar relation. If $\Delta G$ is linearly dependent of temperature $T$ as in protein folding because of the same temperature dependence of conformational energy, then from Eq (1) or Eq (27) one readily obtains

$$\ln W(T) = \frac{S}{T} - RT + \frac{1}{2}\ln T + const.$$

or

$$\ln k_i(T) = \frac{S_i}{T} - R_i T + \frac{1}{2}\ln T + const.$$

(40)

Furthermore, by using (36) (37) or (38) we predict the mean transport time $\ln\frac{1}{t}$ ($\ln\frac{1}{t'}$) obeying the same temperature-dependence law. The non-Arrhenius behavior of the temperature dependence is a feature specific to the quantum conformation change. Through the direct measurement of the temperature dependence of the rate constant or the temperature relation of the mean transport time in a cycle we are able to give deeper insight into the mechanism for the glucose transmembrane transport.

**Perspectives** The binding of a ligand to a membrane receptor results in a conformational change, which then causes a specific programmed response. The present model and theory on sugar transport across membrane can be generalized to other membrane receptors and membrane transport problems. For example, the binding of acetylcholine to an acetylcholine receptor opens a cation channel, the ligand binding at the extracellular side of a G-protein-coupled receptor leads to conformational changes in the cytoplasmic side of the receptor and activate a specific intracellular signalling pathway. A large portion of medications achieve their effect through G-protein–coupled receptors. All these problems are expected to be studied in the framework of the present theory.

Acknowlegement: The author is indebted to Dr Bao YuLai for his help in literature searching, and Drs Lu Jun, Xing Yunqiang and Xue Fudong for their discussions and figure drawing.